\documentclass{article}

\usepackage{epsfig}

\begin{document}
\title{}

\begin{center}
  {\Large\bf   How Does a Dipolar Bose-Einstein Condensate Collapse?}\\[1pc]
  {\large John L. Bohn\footnote{email: bohn@murphy.colorado.edu}, Ryan M. Wilson, and Shai Ronen}\\[1pc]
  {\large \em JILA and Department of Physics, University of Colorado,
    Boulder, CO 80309 USA}\\[1pc]
\end{center}

\vskip 0.5in

\abstract{We emphasize that the macroscopic collapse of a dipolar
Bose-Einstein condensate in a pancake-shaped trap occurs through
local density fluctuations, rather than through a global collapse to
the trap center. This hypothesis is supported by a recent experiment
in a chromium condensate.}

\vskip 1in

\noindent Based on an invited talk in Seminar 6, Physics of Cold
Trapped Atoms, at the 17th International Laser Physics Workshop,
Trondheim, Norway, June 30- July 4, 2008.


\newpage

It is not our purpose in this invited presentation to convince the
reader of the intrinsic interest of a quantum degenerate gas of
dipolar particles.  This task has been admirably accomplished in
several recent reviews \cite{Kurizki04,Baranov08,Lahaye08}, and
continues to play out in laboratories and calculations the world
over. Suffice it to say that these developments are likely to have
an impact on chemistry, condensed matter physics, precision
measurements, and quantum information science. Instead, we are
interested here in a very practical and immediate concern, namely,
the stability of such a gas, and the processes that undermine this
stability.

We will focus here on a Bose-Einstein condensate (BEC) of dipolar
entities, and assume that the dipoles are strictly aligned in a
common direction, which we will identify as the $z$ axis.  In this
circumstance the interaction between dipoles takes the form
\begin{eqnarray}
V({\vec r}) = { d^2 (1 - 3\cos^2 \theta ) \over r^3} + {4 \pi
\hbar^2 a \over m} \delta ( {\vec r})
\end{eqnarray}
where $d$ is the dipole moment, $r$ the distance between dipoles,
and $\theta$ the angle that the intermolecular axis makes with
respect to $z$.  We have included here also an explicit, isotropic
contact interaction that depends on the $s$-wave scattering length
$a$.  The two-body dipolar interaction is attractive for
head-to-tail orientations of the dipoles ($\cos^2 \theta < 1/3$),
and repulsive otherwise.  By contrast, the contact interaction is
equally attractive (if $a<0$) or repulsive (if $a>0$) in all
directions.

In BEC of {\it non-polar} entities, the stability of the condensate
relies on the value of $a$.  When $a>0$, the resulting mean-field
due to interactions accounts for a kind of outward pressure that
inflates the condensate over its non-interacting counterpart; such a
condensate is stable.  When $a<0$, the mean-field energy generates
instead an attractive pressure, which encourages the macroscopic
collapse of the condensate into a high-density lump in the center.
In this high-density environment, the condensate is prone to loss
and heating due to three-body recombination \cite{Dodd96}.  The
mean-field attraction is mitigated somewhat by the zero-point energy
of the trap in which the atoms are held, however.  The result is
that in realistic experiments, the BEC can hold a certain critical
number of atoms before becoming unstable \cite{Dodd96,Bradley97}.
Alternatively, for a fixed atom number, the condensate can be made
to collapse by tuning $a$ to sufficiently negative values
\cite{Cornish00}.

This consideration of repulsive-versus-attractive interactions
raises an interesting issue for the dipolar BEC, since the
dipole-dipole interaction is sometimes attractive, sometimes
repulsive.  In free space without a trapping potential, the
attraction is always enough to de-stabilize the gas.  However, in a
trap, the contribution of the interaction that is dominant depends
on the distribution of the dipoles. To see this, consider the
schematic in Figure 1.  In this figure the solid oval suggests the
shape of the trap confining the dipoles, while the shaded region
represents the resulting shape of the BEC itself.  The trap is
harmonic, with an anisotropy defined by the aspect ratio $\lambda =
\omega_{z} / \omega_{\rho}$ of its azimuthal to radial trap
frequencies.

\begin{figure}
 \caption{Schematic view of the collapse of dipolar BEC. In all cases
 the solid oval represents the trap anisotropy, while the shaded
 gray area represents the anisotropic condensate density. The arrows
 merely suggest the direction of dipole polarization.  In a), a
 nearly spherical trap is unstable against distortion of the BEC into
 a prolate shape,followed by macroscopic collapse.  In b), squeezing
 the trap into a more oblate shape forces the BEC to do the same,
 allowing a greater number of dipoles before collapse occurs.  In c),
 a very oblate trap becomes unstable against local density fluctuations,
 in each of which the condensate is effectively prolate as in a)} \centering
\includegraphics[width=0.9\textwidth] {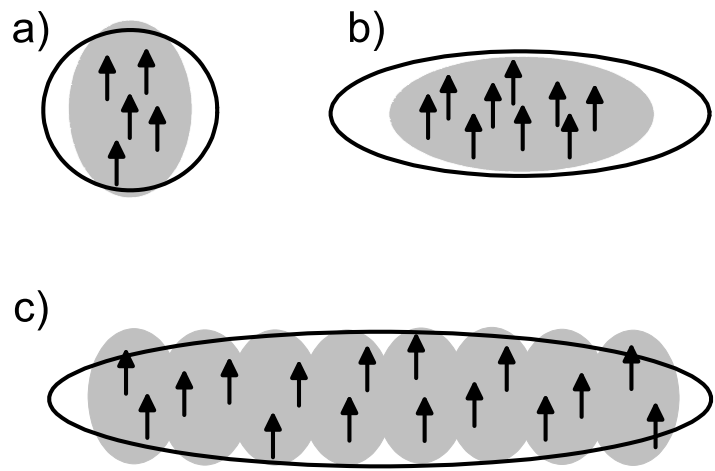}
\end{figure}

Consider Fig. 1a), where the trap is nearly spherical, $\lambda
\approx 1$. Because the dipole-dipole interaction is attractive for
a head-to-tail orientation, the gas can lower its energy by
distorting into a more prolate shape \cite{Yi00,Santos00}. As the
dipole strength grows, this shape becomes more pronounced,
emphasizing the attractive part of the interaction, until at last it
overwhelms the condensate and collapse occurs. In a spherical trap,
and in the absence of a contact interaction, the BEC can stretch to
about 1.6 times higher than it is wide, before collapsing.

An obvious way to overcome this instability is to introduce an extra
energy cost for the cloud to distort into a prolate shape. To do
this, one might contemplate holding the BEC in an oblate trap with
$\lambda \gg 1$, as suggested in Fig. 1b).  In this case, it might
be expected that the gas cannot achieve a prolate shape, regardless
of interaction, because it would have to overcome the immense
harmonic confinement potential in the $z$ direction.  Indeed,
estimates based on a gaussian {\it ansatz} wave function for the
condensate predict that, for $a=0$, the condensate is completely
stable beyond a certain aspect ratio, regardless of the number of
dipoles in the BEC \cite{Santos00}.

Nevertheless, more careful calculations using mean-field theory with
the nonlocal potential (1) \cite{Ronen06} have suggested that the
BEC can become unstable for {\it any} aspect ratio, when the dipole
strength is large enough \cite{Ronen07}. The mechanism by which this
happens is illustrated schematically in Fig. 1c).  In a very prolate
trap, the BEC may reduce its total energy by distorting locally into
small, dense clumps.  In each such clump the attraction can
overwhelm the kinetic energy of trap confinement, just as in the
simpler case in Fig. 1a).

At dipole strengths just below the threshold for instability, these
clumps appear as distortions of the ground state wave function
\cite{Ronen07,Dutta07,Wilson08}. They are ultimately generated by
excited state modes that become soft near the instability, and that
we have dubbed ``roton'' modes for reasons described elsewhere
\cite{Ronen07,Wilson08}. If a just-barely-stable BEC is prepared,
and the interaction strength is suddenly ramped into the unstable
regime, the BEC will exhibit a dynamical collapse that is seeded by
this low-lying roton mode.  The reader can access movies of just
this sort of collapse at
http://grizzly.colorado.edu/$\sim$bohn/movies/collapse.htm. On this
site there are examples of collapse into both radial and angular
excitations. Upon collapsing, the gas can rebound to expand in an
anisotropic pattern, as has been found experimentally and
theoretically \cite{Lahaye08PRL}.

The stability of a dipolar BEC can be probed experimentally, and
indeed this has been done by the Pfau group \cite{Koch08}.  This
experiment produces a BEC of dipolar chromium atoms in traps of
varying aspect ratios. While it is difficult to experimentally
change the atom number or dipole strength on demand, the experiment
can nevertheless change the sign and magnitude of the $s$-wave
scattering length by tuning a magnetic field.  An intrinsically
less-stable BEC requires a more positive scattering length to ward
off its collapse.  In theoretical simulations we can of course
include the scattering length as well.

The stability diagram probed in Ref. \cite{Koch08} is shown in Fig.
2 on a plot of scattering length at which the BEC goes unstable,
versus aspect ratio. In this figure the solid line represents the
boundary between stable (above the line) and unstable (below the
line) BEC, as determined by the softening of the relevant roton
mode.  Consider first the small-aspect ratio limit (left-hand side
of the figure). In this case the trap would readily allow the BEC to
distort in a prolate shape as in Fig. 1a); therefore a large,
positive scattering length is required to stabilize the gas.  By
contrast, on the large-aspect ratio (right-hand) side, this
distortion is suppressed, and a smaller scattering length is
sufficient to stabilize the BEC. Indeed, for the conditions of the
experiment, the BEC can sustain a slightly negative scattering
length and remain stable. This conclusion would no longer hold,
however, for a significantly larger number of chromium atoms
\cite{Wilson09}.

\begin{figure}
\caption{Map of the stability of a dipolar BEC against collapse.
Solid line: stable-unstable boundary as computed by mean-field
theory, determined by the softening of an excited mode to zero
excitation energy.  Dashed line: prediction of the stable-unstable
boundary based on a gaussian {\it ansatz}, which cannot account for
the local collapse mechanism in Fig. 1c).  Data result from the
measurement in Ref. \cite{Koch08}, supporting the mean-field
calculation over the gaussian {\it ansatz}.} \centering
\includegraphics[width=0.9\textwidth]{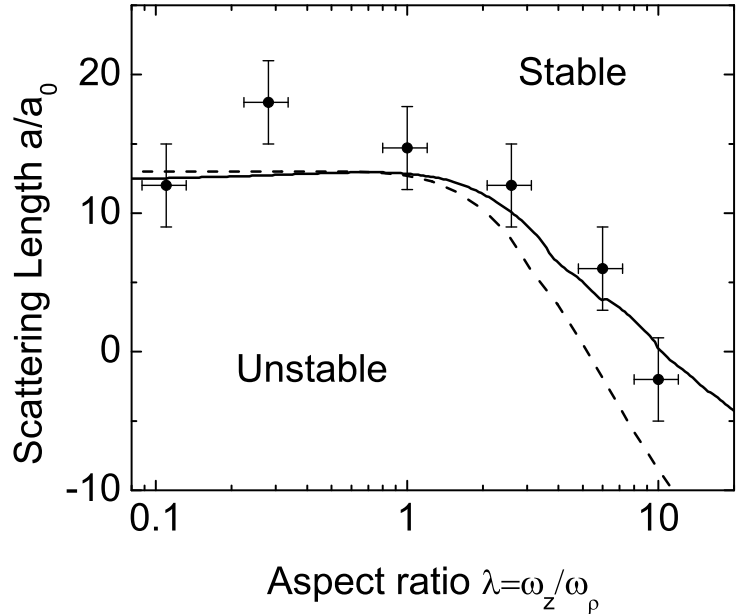}
\end{figure}

The experimental data are also shown in Fig 2.  The agreement with
the mean-field theory is quite good, suggesting that this theory
remains valid for a chromium gas.  In addition, the dashed line
shows the prediction for the stability-instability boundary based on
a gaussian {\it ansatz} wave function.  This approximation is
accurate for small aspect ratio (prolate traps), but suggests
greater stability at high aspect ratio than the mean-field
calculation allows. This is because a gaussian wave function cannot
model local collapse of the kind depicted in Fig. 1c).  The data
therefore suggest that the mean-field description is the correct
one, thus supporting the idea of a collapse due to local density
fluctuations.

This work was supported by the U.S. Department of Energy and the
National Science Foundation.

\end{document}